\documentclass[aps,twocolumn,pr*,amsfonts,showpacs,superscriptaddress,longbibliography]{revtex4-2}
\DeclareUnicodeCharacter{0393}{$\Gamma$}
\usepackage{verbatim,epsfig,amsmath,amssymb,bm,epsf,graphicx,psfrag,bbold,amsthm,amsfonts}
\usepackage[bottom]{footmisc}
\usepackage{hyperref}
\usepackage{cleveref} 
\usepackage{enumitem}
\usepackage{framed}
\usepackage{mathrsfs}
\usepackage{esint}
\setlist{nosep}
\usepackage{placeins}
\usepackage[all]{xy}
\usepackage{color}
\usepackage{xcolor}
\usepackage[utf8]{inputenc}
\usepackage{float}
\usepackage{natbib}
\usepackage{tikz-cd}
\usepackage{verbatim}
\usepackage{leftidx}
\usepackage{physics}
\usepackage{ulem}

\usepackage{pifont}
\usepackage{braket}
\usepackage{booktabs}

\begin{document}
\title{Quantum Electron Quasicrystal}

\author{Pierre-Antoine Graham}
\affiliation{Department of Physics, Massachusetts Institute of Technology, Cambridge, MA-02139, USA}

\author{Filippo Gaggioli}
\affiliation{Department of Physics, Massachusetts Institute of Technology, Cambridge, MA-02139, USA}

\author{Liang Fu}
\affiliation{Department of Physics, Massachusetts Institute of Technology, Cambridge, MA-02139, USA}

\date{\today}

\begin{abstract}
The strongly correlated phases of the homogeneous electron gas constitute the vocabulary of many-body condensed matter physics and find a natural realization in semiconductors.
In this setting, recent neural-network variational Monte Carlo calculations \cite{gaggioliElectronicCrystalsQuasicrystals2025} discovered an unexpected quantum phase of matter in wide quantum wells: an electronic quasicrystal formed by a bilayer Wigner crystals with a $30^\circ$ twist. 
This state defies classical expectations and emerges in a regime dominated by quantum fluctuations.
Here, we develop an analytical framework to reveal its origin.
By computing zero-point energy corrections to bilayer Wigner crystal configurations, we show that quantum fluctuations qualitatively reshape the energetic landscape, destabilizing the classical honeycomb state and selecting the $30^\circ$ quasicrystalline ground state over a broad parameter range.
Our results identify zero-point motion as the mechanism stabilizing the electronic quasicrystal and establish a route to spontaneous moiré physics driven by many-body quantum effects.
\end{abstract}

\maketitle

The competition between energy and entropy gives rise to a rich variety of phases of matter as a function of temperature. 
Atoms arrange into crystalline solids that minimize interaction energy, form liquids when thermal fluctuations dominate, and can even realize quasicrystalline order when competing interactions favor non-periodic structures \cite{Shechtman_1984, Levine_1984, MFMermin1985, PhIcosahedral1985, Widom_1987, Goldman_1993_review}.

In quantum systems, the uncertainty in electron's position leads to a competition between localized and itinerant electron phases. For electrons at zero temperature, quantum fluctuations produce quantum liquids  such as metals, while strong Coulomb interaction drives electron crystals \cite{Andrei2DEGWC, B0WignerShayegan, zhouSignaturesBilayerWigner2021, Smolenski_2021, Li_2021, tsuiDirectObservationMagneticfieldinduced2024c}, first predicted long ago by Wigner \cite{wignerInteractionElectronsMetals1934c}. 
Yet, the nature of quantum melting and possible phases that can emerge in its vicinity remain open questions \cite{BelloLevinShklovskiiEfros1981, PhaseIntermediateKivelson, CoulombFrustatedKivelson, HybridWaintal, falsonCompetingCorrelatedStates2022, KSselfdoping, UnifiedVariationalShiwei}.
In particular, whether there exists quantum analogues of quasicrystals in electronic systems is an outstanding problem.

Semiconductor quantum wells provide a particularly versatile setting in which to explore these questions.
In the two-dimensional limit, the competition between interaction and kinetic energy  gives rise to a transition from a metallic state to a triangular Wigner crystal at low electron densities.
Confinement in the transverse direction introduces an additional degree of freedom, allowing the electron gas to access both monolayer and bilayer regimes \cite{BilayerWignerExpShayegan, zhouSignaturesBilayerWigner2021} and enriching the landscape of competing ground states.
The interplay of 
energy scales
stabilizes a variety of crystalline phases \cite{esfarjaniBilayerWignerCrystal1995, goldoniStabilityDynamicalProperties1996, WignerQuantumBilayer} and offers a controlled platform for probing quantum melting.

Recently, this problem has been revisited using neural-network-based variational Monte Carlo (NN-VMC), which enables unbiased simulations of interacting electrons in continuous space \cite{Pescia_2024, Geier_2025}.
In a previous work \cite{gaggioliElectronicCrystalsQuasicrystals2025}, we uncovered a rich phase diagram of electrons in wide quantum wells, including both monolayer and bilayer Wigner crystals.
Most strikingly, in the bilayer regime near the onset of quantum melting, we discovered a previously unknown quantum phase of matter: an electron quasicrystal formed by a $30^\circ$-twist  between two layers of triangular Wigner crystals.

This quasicrystalline state defies classical expectations. In a purely classical bilayer of point charges, the ground state is known to favor crystalline configurations, such as a honeycomb structure. By contrast, the quasicrystal we identified emerges in a regime where quantum fluctuations are strong and fundamentally alter the energetic hierarchy of competing states. Its stability therefore has no classical analogue and instead originates from zero-point motion of the electrons. This establishes the electronic quasicrystal as a genuinely quantum phase of matter.

In this work, we develop an analytical theory to reveal the quantum origin and stability of the emergent quasicrystal phase in electronic bilayer systems. Our approach is based on quantifying the zero-point energy of collective fluctuations around candidate bilayer configurations, for arbitrary twist angles.

Our analysis reveals that quantum fluctuations qualitatively reshape the phase diagram of bilayer electron systems. While classical energetics favor the honeycomb configuration, we show that sufficiently strong zero-point motion stabilizes twisted electronic structures. Among these, the $30^\circ$ twisted configuration—corresponding to a quasicrystalline arrangement—is found to minimize the total energy and thus emerges as the true ground state in a broad parameter regime. These results provide a transparent physical explanation for the numerical observations of NN-VMC and establish zero-point fluctuations as the key mechanism driving the formation of the electronic quasicrystal.

\begin{figure}
    \centering
    \includegraphics[width=1.0\linewidth]{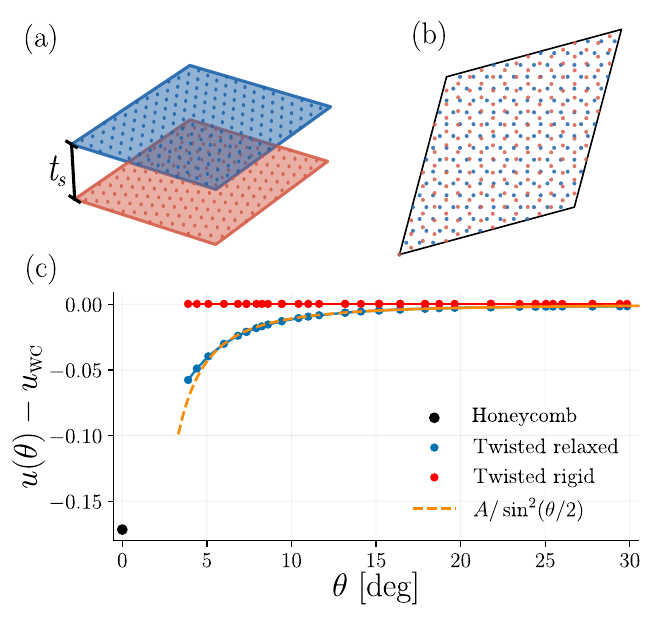}
    \caption{(a) Representation of the bilayer set-up. (b) A $29.84^\circ$-approximant of the quasicrystal. (c) Difference in classical energy  per particle ($10^{-3} E_{\rm H} \cdot r_s^{-1}$) between a twisted configuration at angle $\theta$ and two decoupled triangular layers for $t_s/r_s = 3$. The black point indicates the energy difference for the honeycomb stacking. For two decoupled  Wigner crystal layers $u_{\rm \scriptscriptstyle WC} =  -0.782133\ E_{\rm H} \cdot r_s^{-1}$ \cite{AccurateGSAlves2021}. The analytical angular dependence Eq. \eqref{eq:u_inter_largeTheta} is shown in orange.}
    \label{fig:classical}
\end{figure}

\textit{Classical ground state --} We consider a bilayer system of spin-polarized electrons with density $n_{2D}$ and interparticle distance $r_s = (\pi n_{2D})^{-1/2}$, with two layers $l=1,2$ with separation $t_s$ and of equal density $n_{2D}/2$ as shown in Fig. \ref{fig:classical} (a). Note the difference between $r_s$ and the usual definition involving the density of a single layer given by $r_s^{\rm mono} = \sqrt{2} r_{s}$. Unless stated otherwise, we use the effective Hartree system $\hbar = e = m = a_B = 1$ where $m$ is the effective electron mass and $a_B$ the effective Bohr radius. The effective Hartree energy is $E_{\rm H} = \hbar^2/m a_B^2$.

We denote the position of an electron $i$ in layer $\ell$ as $\mathbf{r}_{i, \ell} = (\boldsymbol{\rho}_{i, \ell}, z_{i, \ell})$ with $z_{i, 1} = -t_s/2$ and $z_{i, 2} =  t_s/2$. The Hamiltonian of Coulomb interacting electrons then reads
\begin{align}\label{eq:H}
 H &= -\frac{1}{2}\sum_{i, \ell}\!\left(\nabla_{i, \ell}^2 + \sum_{\neq i, \ell} \!\frac{1}{\sqrt{|\boldsymbol{\rho}_{i, \ell}\!-\!\boldsymbol{\rho}_{j, \ell'}|^2+\delta_{\ell \neq \ell'}\, t_s^2}}\right)
\end{align}
up to the interaction with a neutralizing background. As we will show later, Equation \eqref{eq:H} captures the essential physics of the bilayer regime that spontaneously occurs in semiconductor wide quantum wells \cite{gaggioliElectronicCrystalsQuasicrystals2025}. 
For large values of $r_s$, interactions dominate over the kinetic energy and the
ground state of the Hamiltonian \eqref{eq:H} is an electron crystal. In particular, for large values of $t_s/r_s$, 
intralayer Coulomb energy dominates over interlayer interaction, and therefore electrons on each layer form a triangular Wigner crystal.

In this work, we study the twist angle $\theta$ dependence of the energy of such bilayer stackings. To start, we focus on the classical Coulomb interaction energy per particle $u$, which scales $\propto 1/r_s$ at fixed $t_s/r_s$. It consists of intralayer terms for each layer, which are minimized by a triangular lattice geometry, and an interlayer term that reads
\begin{align}\label{eq:u_inter}
    u_{\rm inter} 
    = \frac{\pi n_{\rm 2D}}{2} \sum_{\mathbf{G} \neq \mathbf{0}}  \frac{ e^{- |\mathbf{G}| t_s} }{|\mathbf{G}|} \rho_1(\mathbf{G}) \rho_{2}(-\mathbf{G}),
\end{align}
where $\rho_\ell(\mathbf{G}) = \frac{2}{N}\sum_{i} e^{i \mathbf{G} \cdot \boldsymbol{\rho}_{i, \ell}}$ the Fourier components of the charge density in layer $\ell$ for $N$ electrons.  
Equation \eqref{eq:u_inter} favors commensurate stackings: the product $\rho_{\ell}(\mathbf{G})\rho_{\ell'}(-\mathbf{G}) \propto e^{i \mathbf{G} \cdot \mathbf{s}}$ is generically non-zero, and since $\int \text{d}^2\mathbf{s}\, \, u_{\rm inter}(\mathbf{s}) = 0$ there must exist a shift $\mathbf{s}$ that lowers energy relative to the decoupled limit value $u_{\rm inter}  = 0$. On the other hand, $u_{\rm inter}$ vanishes if there is no common Fourier density components. 

For a rigid bilayer at sufficiently large $t_s/r_s$, the honeycomb configuration ($\theta = 0$) minimizes the interlayer energy \eqref{eq:u_inter}, which reads
\begin{equation}\label{eq:gap_hc}
u_\text{inter}\sim - \pi \,n_{2D} \,e^{-Gt_s}/G,
\end{equation}
with $G = 4\pi/\sqrt{3}a \sim r_s^{-1}$ and $a$ the lattice constant of the triangular layers.
For \textit{twisted rigid} bilayers, on the other hand, we find that the interlayer energy is negligible even for commensurate stackings, as shown by the red line in Fig.~\ref{fig:classical} (c).

\begin{figure*}[t] 
    \centering
    {\label{fig:panel_a}\includegraphics[width=1\textwidth]{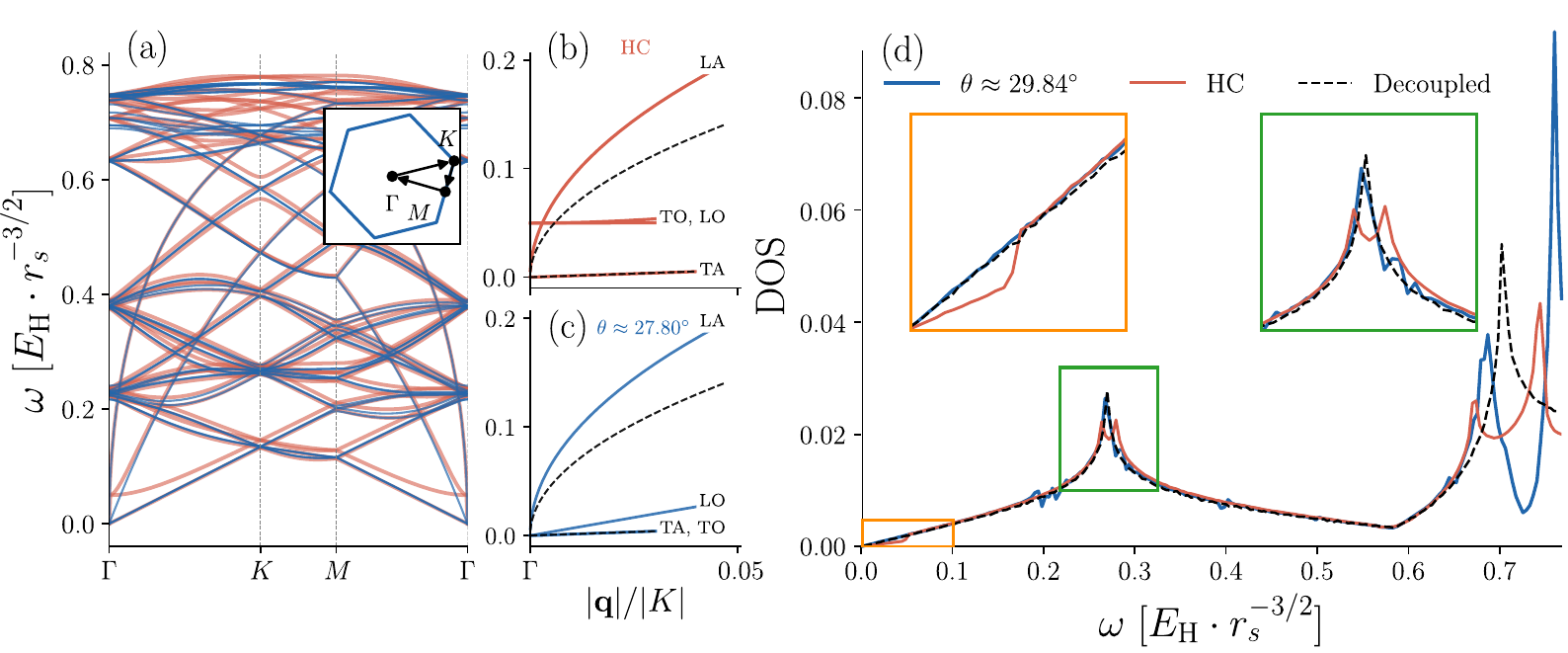}}
    \caption{(a) Phonon spectrum of the quasicrystal approximant $\theta \approx 27.80^\circ$ (blue) and of the honeycomb stacking (red), after folding back to the mini-Brillouin zone of the approximant for direct comparison. Bilayer thickness $t_s/r_s = 3$. (b), (c) Zoom-in on the low-lying modes for the honeycomb and the quasicrystal approximant. Dashed lines show the degenerate phonon and plasmon dispersions for two decoupled layers. (d) Density of states per particle of the quasicrystal approximant $\theta \approx 29.84^\circ$, honeycomb stacking and $t_s/r_s \to \infty$ decoupled layers. The orange inset shows the enhanced density of states at low frequency for quasicrystal vs honeycomb, while the green inset zooms-in on the first peak.}
    \label{fig:Band_DOS}
\end{figure*}

In a realistic bilayer, however, electrons in each layer experience interlayer forces and relax toward a (local) energy minimum that deviates from the rigid triangular lattice. 
In order to obtain mechanically stable configurations at different twist angles, we start from two triangular layers and use gradient descent to minimize the classical Coulomb energy. 

Similarly to other twisted bilayers \cite{namLatticeRelaxationEnergy2017, ezziAnalyticalModelAtomic2024, PhysRevB.99.205134}, relaxation creates a modulation with characteristic lengthscale $L_M = \frac{a}{2\sin(\theta/2)}\sim|\mathbf{G}_1 - \mathbf{G}_2|^{-1}$ that imprints common $\rho_l(\mathbf{G}_{l'})$ components in each layer, thus lowering the interlayer energy \eqref{eq:u_inter}. 
This comes at the expenses of an intralayer elastic energy cost $\kappa (\delta r/L_M)^2$, with $\delta r$ being the typical displacement and $\kappa \sim a^{-1}$ the relevant elastic constant.
At small angles, $L_M\gg 1$ and the elastic energy cost is negligible, thus leading to strong relaxation.
On the other hand, the  elastic energy becomes  comparable to $u_{\rm inter}\sim n_{2D}e^{-G t_s}/G$ for twist angles 
\begin{equation}
\sin^2(\theta/2) \gtrsim e^{-G t_s}/\kappa\, a \propto e^{-G t_s},
\end{equation}
beyond which the effect of relaxation is weak.
Balancing the elastic and interlayer forces in the large angle limit, $\kappa\,\delta r/L_M^2 \sim e^{-G t_s} / a^2$, we then find typical displacements of order  
$\delta r\sim L_M^2 e^{-G t_s}/a$
and, correspondingly, the classical energy is lowered by
\begin{equation}\label{eq:u_inter_largeTheta}
 u(\theta) \sim  e^{-2Gt_s}/a\sin^2(\theta/2),
 \end{equation}
as detailed in  Section~\ref{sec:angle_dep} of the SM~\cite{supplementary}.

The classical energies after relaxation are shown in blue in Fig.~\ref{fig:classical} (c) for $t_s/r_s = 3$. These energies were using the adapted Ewald summation method \cite{grzybowskiEwaldSummationElectrostatic2000a, goldoniStabilityDynamicalProperties1996} obtained for commensurate twist angles where the moiré pattern exactly repeats as a superlattice (see Sec. \ref{sec:CS} of the SM \cite{supplementary} for more details) -- 
the smoothness of the angular dependence then allows to extrapolate to the neighboring incommensurate configurations. In particular, we can always find a family of commensurate approximants that converge towards a given incommensurate stacking.
Comparing to the numerics, we find that our analytical scaling \eqref{eq:u_inter_largeTheta} (orange dashed) 
reproduces the classical energy $u(\theta)$ over a wide range of twist angles angles, for sufficiently large values of $t_s/r_s$.

At the classical level, we have therefore found that relaxation plays an important role, but 
makes the honeycomb stacking and quasicrystal stackings as the lowest and highest energy configurations, respectively.
This picture will change upon including the effect of quantum fluctuations, as we will now show below.

\textit{Zero-point energy --} 
Quantum fluctuations modify the stability of classical configurations by means of the zero-point energy  contribution $\text{ZPE} = \sum_i\hbar\omega_i/2$ from their collective excitations.
To calculate the ZPE as a function of twist angle, we consider small displacements of an electron $\alpha,l$ around the classical relaxed configuration,
\begin{align}
\mathbf{u}_{\mathbf{R}\alpha \ell} = \frac{1}{\sqrt{N_{\rm cell}}} \sum_{\mathbf{q}} \boldsymbol{\gamma}_{\mathbf{q} \alpha \ell} \,e^{i \mathbf{q}\cdot \mathbf{R}}
\label{eq:disp},
\end{align}
where $\mathbf{R}$ is the superlattice site and $N_\text{cell}$ is the number of repeating supercells in the bilayer stacking \cite{supplementary}. The wave-vector $\mathbf{q}$ belongs to the mini-Brillouin zone of the superlattice, which reduces to a single point $\mathbf{q} = 0$ in the case of an incommensurate stacking formed by a single, infinitely large, supercell.

Rewriting the Hamiltonian \eqref{eq:H} in terms of the displacements $\mathbf{u}_{\mathbf{R}\alpha \ell}$ and expanding to quadratic order in $\boldsymbol{\gamma}_{\mathbf{q} \alpha \ell}$, we
obtain the coupled-harmonic oscillator problem
\begin{align}
    H = -\frac{1}{2} &\sum_{\mathbf{R} \alpha \ell }\nabla_{\mathbf{R} \alpha \ell}^2 
     +   \frac{1}{2} \sum_{\mathbf{q}\alpha\ell \beta \ell'}   \boldsymbol{\gamma}_{\mathbf{q} \beta \ell'}^\dagger  \mathbf{D}_{\alpha\beta}^{\ell \ell'}(\mathbf{q}) \boldsymbol{\gamma}_{\mathbf{q} \alpha \ell},
\end{align}
where $\mathbf{D}_{\alpha \beta}^{\ell \ell'}(\mathbf q)$ are components of the dynamical matrix and $-i \nabla_{\mathbf{R}\alpha\ell}$ is conjugate to $\mathbf{u}_{\mathbf{R}\alpha \ell}$. 

To solve this problem exactly, we proceed to finding the normal modes $\boldsymbol{\gamma}_{\mathbf{q} \alpha \ell}$ that diagonalize $\mathbf{D}_{\alpha \beta}^{\ell\ell'}(\mathbf{q})$ for all $\mathbf{q}$ in the mini-Brillouin zone.  
The associated eigenvalue spectrum $\lambda_{m}(\mathbf{q})$ with band index $m = 1, \cdots, 2N$ will then determine the normal mode frequencies as $\omega_{m}(\mathbf{q}) = \sqrt{\lambda_{m}(\mathbf{q})}$. Because the dynamical matrix describes the curvature of the Coulomb potential, scaling as $1/r_s^3$, it follows that $\omega\propto r_s^{-3/2}$ at fixed $t_s/r_s$.  

To begin, we consider the simple limit $t_s/r_s \to \infty$ of two decoupled Wigner crystals. 
In this case, the spectrum contains two degenerate copies of the transverse phonon- and longitudinal plasmon modes characteristic of the monolayer. 
As $t_s/r_s$ is reduced, the interlayer Coulomb interaction lifts the degeneracy and splits them into two pairs of acoustic (in-phase) and optical (out-of-phase motion of the layers) modes. 
Independently of the twist angle, both the transverse (TA) and longitudinal (LA) acoustic modes remain gapless, due to the translation symmetry of Eq.\ \eqref{eq:H}. On the other hand, the twist-angle dependence of $u_\text{inter}$ crucially determines the gap of the optical modes via the dispersion relation $\omega(\mathbf{q};\theta)\sim \sqrt{\omega_{\scriptscriptstyle \text{  WC}}^2(\mathbf{q}) + |u_\text{inter}(\theta)|}$, both transverse (TO) and longitudinal (LO): the $\mathbf{q} = 0$ gap will range from $\sim e^{-Gt_s/2}$ for the optimally stacked honeycomb bilayer (see Eq.~\eqref{eq:gap_hc}), to an exponentially smaller gap $\sim e^{-Gt_s}$ (see Eq.\ \eqref{eq:u_inter_largeTheta}) for the $30^\circ$-twisted quasicrystal, where interactions between the layers are weakest.    

Figure \ref{fig:Band_DOS} $(a)$-$(c)$ shows the normal mode spectrum for the honeycomb bilayer (red) and quasicrystal approximant (blue), after folding the honeycomb bands into the same mini-Brillouin zone for direct comparison. The insets provide zoom-ins of the dispersion at small $|\mathbf{q}|$, highlighting the different gap-size, with the monolayer curves (black dashed) serving as a reference. 

As anticipated, all four modes remain (almost) gapless in the quasicrystal state.
This is a consequence of an emergent non-local continuous relative-shift symmetry: up to exponentially small relaxation effects, shifting one quasicrystal layer relative to the other modifies local patterns but leaves the total energy unchanged, preventing TO and LO modes from developing a gap in a rigid quasicrystal. At small $\mathbf{q}$, the gapless optical modes correspond to a nearly uniform relative displacement of the two layers, with a vanishing energy cost as $\mathbf{q} \to \mathbf{0}$.
Protected by the relative-shift symmetry, these modes are ``phasons'' \cite{Phason0}. Unlike regular phonons, the phasons propagate in a diffusive manner, owing to the non-local nature of the shift symmetry \cite{PhasonsZeyher1982, Qiang2022}.

A useful measure of the distribution of collective modes is provided by the density of states $\text{DOS}(\omega)$ per particle, which quantifies the number of normal modes in a small frequency range centered around $\omega$.
This is shown in Fig.~\ref{fig:Band_DOS} $(d)$ for the case of a honeycomb (red) and large quasicrystal approximant with $\theta \approx 29.84^\circ$ (blue).
Evidently, the two peaks present in the DOS of decoupled Wigner crystal layers (black dashed) split differently for the two stackings: in the quasicrystal, more weight is carried to lower frequencies and modes up to the first peak closely match the decoupled layers density of states, while in the honeycomb more DOS is shifted to higher energies.

\begin{figure}
    \centering
    \includegraphics[width=1\linewidth]{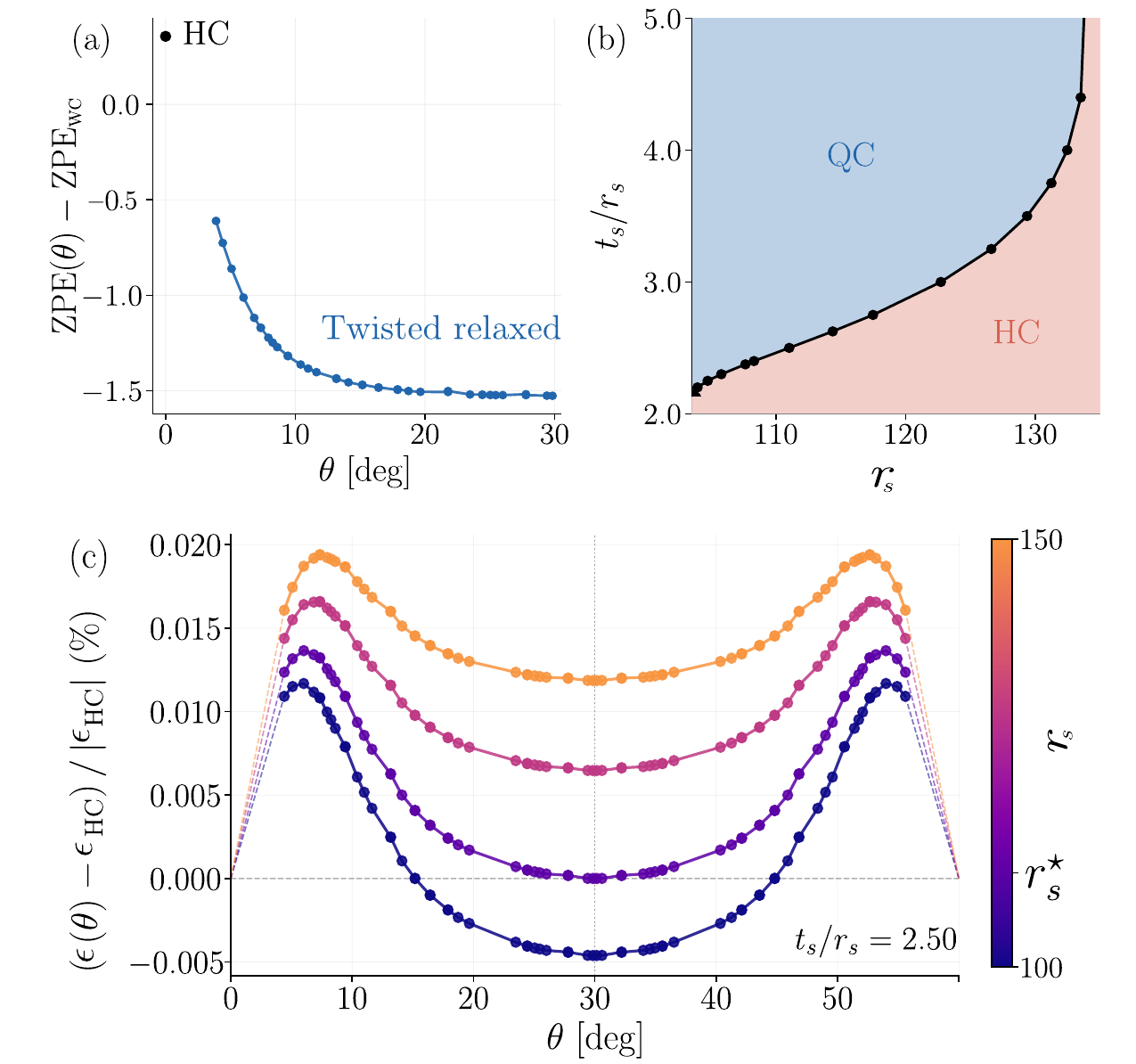}
    \caption{$(a)$ Difference in zero point energy  per particle ($10^{-3} E_{\rm H} \cdot r_s^{-3/2}$) between twisted configuration at angle $\theta$ and decoupled triangular layers when $t_s/r_s = 3$. For two decoupled layers, $\text{ZPE} =  0.483821\ E_{\rm H}\cdot r_s^{-3/2}$ \cite{AccurateGSAlves2021}. $(b)$ First order phase transition line $r_s^*$ separating the honeycomb and quasicrystal states, as obtained by setting $\epsilon(\theta) = \epsilon_{\rm \scriptscriptstyle HC}$ in Eq.~\eqref{eq:epsilon_theta}. $(c)$ Angular dependence of the total energy per particle, defined relative to honeycomb state, as a function of the twist angle $\theta$ for different values of $r_s$ at $t_s/r_s = 2.5$.}
    \label{fig:ZPE}
\end{figure}

From the density of states, the zero-point energy per particle of a bilayer stacking with twist angle $\theta$ can be compactly written as
\begin{align}
    \text{ZPE}(\theta) = \frac{1}{2} \int \text{d}\omega \ \omega \ \text{DOS}(\omega;\theta). 
\end{align}
Fig. \ref{fig:ZPE} (a) gives the zero point energy per particle for different twists angles at $t_s/r_s = 3$, defined with respect to the value $\rm ZPE_{\rm \scriptscriptstyle WC}$ for two decoupled layers. As was the case for the the classical energy shown in Fig.\ \ref{fig:classical} (c), the quantum correction has a weaker dependence on the angle $\theta$ at larger twists, where the interlayer coupling is small. Crucially, however, the twist angle dependence of the zero-point energy  displays the opposite trend: it is minimal for the $30^\circ$-quasicrystal and grows larger with the twist angle, reaching its maximum with the honeycomb stacking.

 With the zero point energy at hand, we are ready to study the total energy $\epsilon(\theta)$ of the bilayers as a function of the twist angle, and determine their relative stability -- this will ultimately depend on the parameter $r_s$, which sets the scale of quantum fluctuations. 

We make the $r_s$ dependence explicit by rewriting $u = \tilde{u}/r_s$ and $\text{ZPE} = \tilde{z}/ r_s^3$ where $\tilde{u}(t_s/r_s
), \tilde{z}(t_s/r_s
)$ are now functions of the classical parameter $t_s/r_s$ only. 
The total energy per particle at twist angle $\theta$, defined relative to honeycomb stacking, then reads 
\begin{align}\label{eq:epsilon_theta}
        \epsilon(\theta) - \epsilon_{\rm \scriptscriptstyle HC} &\approx \frac{\tilde{u}}{r_s} + \frac{\tilde{z}}{r_s^{3/2}} + O(r_s^{-2}),
\end{align}
or equivalently, after restoring dimensional factors ($a_B =  4\pi\varepsilon \hbar^2/ m e^2$ is the effective Bohr radius),
\begin{align}\label{eq:epsilon_theta_real_units}
   E(\theta) - E_{\rm HC} &\approx \frac{e^2 (\pi n_{\rm 2D})^{\frac{1}{2}}}{4\pi\varepsilon}\left[\tilde{u} + \sqrt{a_B(\pi n_{\rm 2D})^{\frac{1}{2}}} \,\tilde{z}\right],
\end{align}
with $\tilde{u}(t_s/r_s) > 0$ and $\tilde{z}(t_s/r_s) < 0$. 
At very large $r_s$, the classical term dominates, and the honeycomb stacking is favored. On the contrary, as $r_s$ decreases, the zero point energy term dominates Eq.\ \eqref{eq:epsilon_theta} and $\epsilon(\theta) < \epsilon_{\rm \scriptscriptstyle HC}$ with the lowest energy associated with the lowest ZPE found at $30^\circ$.

The first-order phase transition line $\epsilon(30^\circ) =\epsilon_{\rm \scriptscriptstyle HC}$ is shown in black in Fig.~\ref{fig:ZPE}$\,(b)$ as a function of $r_s$ and $t_s/r_s$. The angular dependence of the total energy is displayed in Fig.~\ref{fig:ZPE}$\,(c)
$, and shows that the $30^\circ$-quasicrystal goes from local- to global minimum upon decreasing $r_s$.
More details of the phase transition are reported in the supplementary \cite{supplementary}.
We have therefore found that the quasicrystal becomes the true ground state when quantum fluctuations are sufficiently large, confirming the 
genuinely quantum mechanical nature of this state.

\begin{figure}
    \centering
    \includegraphics[width=0.95\linewidth]{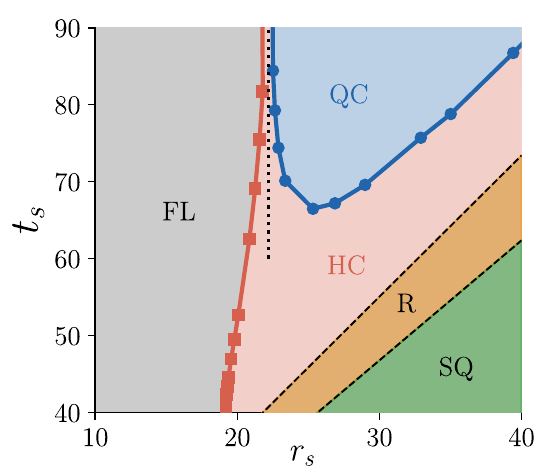}
    \label{fig:phase}
    \caption{Small-$r_s$ phase diagram showing the Lindemann-criterion melting lines for the quasicrystal (circles) and honeycomb (squares) states, as a function of $r_s$ and $t_s/r_s$. The vertical dotted line represents the melting line for a single Wigner crystal layer. The gray region represents the Fermi liquid. We indicate regions where the classical ground state is a staggered rhombic stacking (R) and staggered square stacking (SQ) \cite{goldoniStabilityDynamicalProperties1996} with yellow and green colors, respectively.}
\end{figure}   

\textit{Lindemann criterion -- } The above analysis relies on the harmonic approximation, which becomes invalid as fluctuations increase and eventually melt the crystal stacking into a bilayer Fermi liquid. This calls for an analysis of the Lindemann melting criterion \cite{goldoniStabilityDynamicalProperties1996, esfarjaniBilayerWignerCrystal1995}, which defines the ordered phase from the condition that average fluctuations cannot exceed a certain fraction of the lattice spacing, i.e.,
\begin{align}
    \gamma_{\scriptscriptstyle L}  = \max_{\alpha} \sqrt{\langle |\mathbf{u}_{\mathbf{R}\alpha}|^2/a^2  \rangle} < 0.3,
\end{align}
where the value $0.3$ is obtained from imposing that the monolayer Wigner crystal melts at $r_s^{\rm mono} \approx 31$ (correspondingly $r_s\approx 22$), as found by QMC on spin-polarized 2D electron gas \cite{QMC2009Drummond, QMC2024Drummond}.
This analysis is of particular importance in the case of the $30^\circ$-quasicrystal, where the weak interlayer coupling and the presence of soft, gapless optical modes conjure to destabilize the crystalline phase.

Figure \ref{fig:phase} shows the melting line for the honeycomb and quasicrystal phases as a function of $r_s$ and $t_s$. For an honeycomb stacking, zero-point fluctuations become more energetically expensive as the layers are brought closer to another, shifting the melting line \cite{esfarjaniBilayerWignerCrystal1995} to smaller values of $r_s$ than for decoupled layers (vertical dotted). On the opposite, the quasicrystal stacking saturates the Lindemann criterion already at higher values of $r_s$, i.e., for weaker quantum fluctuations. At low $t_s/r_s$, we complete the phase diagram by indicating the classically favored rhombic stacking (orange) and square stacking (green) phases \cite{goldoniStabilityDynamicalProperties1996} replacing the honeycomb stacking. 

Importantly, comparison of Fig.~\ref{fig:ZPE}$\,(b)$ and Fig.~\ref{fig:phase} shows that there exists a wide parameter range where the quasicrystal state remains stable and energetically favorable. Finally, we note that the phase diagram in Fig.~\ref{fig:phase} is in  good quantitative agreement with the NN-VMC results reported in Ref.~\cite{gaggioliElectronicCrystalsQuasicrystals2025}.

\textit{Discussion --} The strongly correlated phases of the homogeneous electron gas constitute the vocabulary of many-body condensed matter physics \cite{Perdew_2003}. In this work, we have analyzed a particular instance of this model -- the electron bilayer -- and determined its diagram as a function of electron density and layer separation. We identify a broad region of parameter space in which ground state is the electron quasicrystal, and elucidate its quantum mechanical origin.
Remarkably, the resulting phase diagram is in excellent agreement with the result of unbiased NN-VMC simulations reported in Ref.~\cite{gaggioliElectronicCrystalsQuasicrystals2025} for wide quantum wells, confirming the robustness of our conclusions.

The electronic quasicrystal has no classical analogue: it is stabilized by zero-point motion and disappears in the classical limit. This mechanism is generic and suggests that quantum quasicrystals may arise more broadly in systems where competing classical configurations are close in energy, and quantum fluctuations lift the degeneracy.

Since the quantum quasicrystal studied here emerges spontaneously from  Coulomb interactions,
it is directly accessible in various solid-state platforms, including semiconductor quantum wells, 
TMD heterostructures and rhombohedral graphene, where signatures of electron crystals have been reported. 
Its presence could be detected through transport signatures \cite{Deng_2016}, optical probes of excitonic umklapp scattering \cite{Smolenski_2021, zhouSignaturesBilayerWigner2021}, or direct imaging using scanning tunneling microscopy \cite{Li_2021, tsuiDirectObservationMagneticfieldinduced2024c}.

Our results connect the rich phenomenology of quasicrystals to the conceptual simplicity of the electron gas, revealing a new surprise in one of the oldest systems in condensed matter physics.
Looking ahead, this invites us to think about  related novel strongly correlated electronic phases.
One intriguing possibility is the formation of a nematic bilayer Fermi liquid with dodecagonal rotational symmetry \cite{nematicClassification2024, nematicTwistedBilayers2024}, driven by incommensuration in the vicinity of the quasicrystal melting line.
Finally, an interesting direction is the extension of the quantum fluctuation mechanism to bosonic systems, where quantum quasicrystals have been previously proposed for ultracold atom systems with Rashba spin-orbit coupling \cite{QuantumQCSOC} and engineered long-ranged interactions \cite{Quantumcluster2020, Mendoza2022}.

\textit{Acknowledgments --} 
This work was supported by the National Science Foundation (NSF) Convergence Accelerator Award No. 2235945. 
F.G. is grateful for the financial support from the Swiss National Science Foundation (Postdoc.Mobility Grant No. 222230).  
L.F. was supported by a Simons Investigator Award from the Simons Foundation.  The authors acknowledge the
MIT SuperCloud and Lincoln Laboratory Supercomputing Center for providing resources \cite{reutherInteractiveSupercomputing400002018b}.

\bibliography{biblio.bib}

\newpage 
\onecolumngrid
\newpage
\makeatletter

\begin{center}
\textbf{\large Supplementary materials for:\\ Quantum Electron Quasicrystal}
\\[10pt]
Pierre-Antoine Graham$^{1}$, Filippo Gaggioli$^{1}$, Liang Fu$^{1}$\\
\textit{$^1$Department of Physics, Massachusetts Institute of Technology, Cambridge, MA-02139, USA}
\end{center}
\vspace{20pt}

\setcounter{figure}{0}
\setcounter{section}{0}
\setcounter{equation}{0}

\renewcommand{\thefigure}{S\@arabic\c@figure}
\makeatother

\section{Commensurate Triangular Stackings}
\label{sec:CS}

To obtain the angle dependence of the classical energy and the normal mode spectrum, we use commensurate bilayer twisted stackings of triangular lattices. Here, we illustrate the details of the construction \cite{pleasantsPlanarCoincidencesNfold1996, meleCommensurationInterlayerCoherence2010a, lopesdossantosContinuumModelTwisted2012} of the minimal supercell of such stackings. 

In layer $\ell$, we place a triangular lattice with lattice constant $a$, unit cell area $A = \sqrt{3}a^2/2$, and primitive vectors $\mathbf{a}_{1}^{\ell}, \mathbf{a}_{2}^{\ell}$. The stacking is specified by a rotation by an angle $\theta$ of $\ell = 2$ with respect to $\ell = 1$. For a commensurate twist angle $\theta \in [0, 60^\circ]$, the two lattices share special coincidence points forming a superlattice. The associated supercell gives the smallest repeating unit of the bilayer crystal pattern and is generated by primitive vectors $\mathbf{A}_1, \mathbf{A}_2$ that connect neighboring coincidence points.

To construct  $\mathbf{A}_1, \mathbf{A}_2$, we use a complex representation of the monolayer triangular lattices. In layer $1$, we choose $\mathbf{a}_1^1 = 1$,  $\mathbf{a}_2^1 = e^{i \pi/3}$ such that a general lattice point of this layer is $z_1 = p + q e^{i \pi/3}$ for integers $p, q$. In parallel, points $z_2$ of layer $2$ are obtained by $\theta$ rotation as $z_2 = z_1 e^{i \theta}$. It follows that the coincidence condition $z_1' = z_2$ for the two layers can be recast in terms of points of layer $1$ as $e^{i \theta} = z_1/z_1'$.

Given that $\theta$ is a \textit{commensurate twist}, a solution with minimal length $|z_1|=|z_1'|$ is provided by the unique factorization property of the Eisenstein integers \cite{pleasantsPlanarCoincidencesNfold1996} as $z_1 = Z$ and $z_1' = Z^\star$ where $Z = p + q e^{i \pi/3}$ for the coprime integers $(p, q)$ and $p \neq q \ \text{mod} \ 3$ \cite{meleCommensurationInterlayerCoherence2010a}. 
We have therefore found that, at commensurate twist angle $\theta$, the first superlattice primitive vector is given by
\begin{align}
    \cos(\theta)  
    = \text{Re}\left(\frac{Z}{Z^\star}\right) = \frac{p^2 + pq - q^2/2}{p^2 + pq + q^2}\label{eq:angle}.
\end{align}
To find a pair of them, we simply have to use that  rotations of both layers by $e^{i \pi/3}$ leave the full stacking unchanged, which yields that $Z e^{i \pi/3}$ is also a superlattice point (with the same minimal length as $Z$). The supercell primitive vectors therefore reads
\begin{align}
    &\mathbf{A}_1 = Z = p \mathbf{a}_{1}^{1} + q \mathbf{a}_2^{1}, \quad \mathbf{A}_2 = e^{i \pi/3} Z 
    = - q \mathbf{a}_1^{1} + (p + q) \mathbf{a}_2^{1}.
\end{align}
The basis pattern of the supercell spanned by $\mathbf{A}_{1,2}$ is specified by $N$ points where 
\begin{align}
    N = 2\frac{A_{\rm s.c.}}{A}
    = 2 (p^2 + pq + q^2) 
\end{align}
and density is $n_{\rm 2D} = N/A_{\rm s.c.}$. Incommensurate angles such as $30^\circ$ are $N \to \infty$ limits of a $(p, q)$ sequence of approximants. The first five angles and associated $(p, q)$ pairs in the $30^\circ$ approximant sequence are listed in Table \ref{tab:placeholder}.

In reciprocal space, the superlattice generated by $\mathbf{A}_1, \mathbf{A}_2$ has primitive vectors 
\begin{align*}
    \mathbf{B}_1 
    = \frac{2\pi}{A_{\rm s.c.}} \hat{z} \times \mathbf{A}_2,\quad \mathbf{B}_2 
    = -\frac{2\pi}{A_{\rm s.c.}} \hat{z} \times \mathbf{A}_1 
\end{align*}
In a crystal with the periodicity of the superlattice, the phonon modes have wave vectors $\mathbf{q}$ within the mini-Brillouin zone of this reciprocal lattice.

An interesting dual problem relevant to the analysis of Section \ref{sec:angle_dep} consists in finding the coincidence points of the twisted reciprocal lattices of each layer. The reciprocal lattice associated with layer $\ell$ has primitive vectors 
\begin{align}
  \mathbf{b}_{1}^\ell = \frac{2\pi}{A} \hat{z} \times \mathbf{a}_2^\ell, \quad \mathbf{b}_{2}^\ell = -\frac{2\pi}{A} \hat{z} \times \mathbf{a}_1^\ell  
\end{align}
with norm $G = 4\pi/\sqrt{3}a$. Adapting the above results, the coincidence points of reciprocal lattices $1, 2$ form a super- reciprocal lattice generated by primitive vectors $\mathbf{B}_1^{\rm r}$, $\mathbf{B}_2^{\rm r}$. When layer $1$ and $2$ are twisted by an angle $\theta$, the reciprocal lattices are also twisted by $\theta$ corresponding to the same pair of integers $(p, q)$ as the direct lattice. It follows that $|\mathbf{B}_1^{\rm r}|^2 = |\mathbf{B}_2^{\rm r}|^2 = G^2(p^2 + pq + q^2) = G^2 N/2$ are inversely proportional to $|\mathbf{B}_1|^2 = |\mathbf{B}_2|^2 = 2G^2/N$.

\begin{table}[h!]
    \centering
    \setlength{\tabcolsep}{8.5pt} 
    \begin{tabular}{c|ccccc}
    $(p,q)$  & $(2,1)$ & $(3,1)$ & $(8,3)$ & $(11,4)$ & $(30,11)$ \\
    \hline
    $N$      & $14$    & $26$    & $194$   & $362$    & $2702$    \\
    Twist & $21.79$ & $27.80$ & $29.41$ & $29.84$  & $29.96$   \\
    \end{tabular}
    \caption{First commensurate stackings in the optimal sequence converging to a $30^{\circ}$ twisted triangular bilayer. The value of the twist shown is $\text{min}(\pi/6 - \theta, \theta)$ with $\theta$ from Eq.~\eqref{eq:angle}.}
    \label{tab:placeholder}
\end{table}

\section{Ewald Summation}
\label{sec:S_ewald}

For a sequence of commensurate angles with $N \leq 402$, we relax the rigid triangular stacking supercells in periodic boundary conditions following the gradient of the Coulomb potential. The equilibrium configuration of electron $\alpha$ in layer $\ell$ is specified by $\mathbf{r}_{\alpha, \ell} = (\boldsymbol{\rho}_{\alpha, \ell}, z_{\alpha, \ell})$ and the associated value of the potential per unit cell is reported in Fig. \ref{fig:classical} (c). In a commensurate stacking with $N_{\rm s. cell}$ copies of the supercell in periodic boundary conditions, we consider small displacements from equilibrium of the form given by Eq.~\eqref{eq:disp}: plane waves over the superlattice sites $\mathbf{R}$ with wave vector $\mathbf{q}$ in the mini-Brillouin zone of the superlattice, modulated by an internal displacement pattern $\boldsymbol{\gamma}_{\alpha \ell \mathbf{q}}$ that repeats from one supercell to the next. The residual superlattice translation symmetry decouples modes in distinct $\mathbf{q}$ sectors, so the normal-mode spectrum shown in Fig.~\ref{fig:Band_DOS}(a) is obtained by diagonalizing the $2N \times 2N$ dynamical matrix at each $\mathbf{q}$. 

Both the classical equilibrium configuration and the dynamical matrix require lattice sums of the long-range Coulomb interaction, which we evaluate using Ewald summation. In this section, we give the Ewald summation for the Coulomb potential and dynamical matrix elements. The rely on 
\begin{align}
S(\mathbf q, \mathbf{r})
=
\sideset{}{'} \sum_{\mathbf R}
\frac{e^{i\mathbf q\cdot\mathbf R}}{\left|\mathbf R+\mathbf{r}\right|}
\end{align}
where $\mathbf{R}$ ranges over superlattice sites and $\mathbf{r} = (\boldsymbol{\rho}, z)$. The prime sum $\sum^{'}$ indicates we exclude $\mathbf{R} = \mathbf{0}$ when $\mathbf{r} = \mathbf{0}$. Following \cite{grzybowskiEwaldSummationElectrostatic2000a, goldoniStabilityDynamicalProperties1996}, we split $S(\mathbf{q}, \mathbf{r})$ into direct and reciprocal components $S = S_{\rm dir.} + S_{\rm rec.}$:
\begin{align}    
&S_{\rm dir.}(\mathbf q, \mathbf{r}) = \frac{1}{\sqrt{\pi}}\sideset{}{'}\sum_{\mathbf{R}} e^{ i\mathbf q\cdot\mathbf{R}}
\int_{\gamma}^{\infty}t^{-1/2}e^{-tz^2}
e^{-t|\mathbf{R} + \boldsymbol{\rho}|^2}\,dt,\\
&S_{\rm rec.}(\mathbf q, \mathbf{r}) = \frac{1}{\sqrt{\pi}}\sum_{\mathbf{R}} e^{i\mathbf q\cdot\mathbf{R}}
\int_{0}^{\gamma}t^{-1/2}e^{-tz^2}
e^{-t|\mathbf{R} + \boldsymbol{\rho}|^2}\,dt - \delta_{\mathbf{r}, \mathbf{0}}\frac{1}{\sqrt{\pi}} 
\int_{0}^{\gamma}t^{-1/2}
\,dt.
\end{align}
The splitting is controlled by $\gamma$ and we used the efficient value $\gamma = (2.4/\sqrt{A_{\rm s.c}})^2$ \cite{CASINO2019manual}. 
While $S_{\rm dir.}$ is evaluated in direct space, $S_{\rm rec.}$ is converted to a sum on the reciprocal lattice vectors $\mathbf{G}$ of the superlattice by Poisson summation. We have  
\begin{align}
&S_{\rm dir.}(\mathbf q, \mathbf{r})
=
\frac{1}{\sqrt{\pi}} \sideset{}{'} \sum_{\mathbf{R}} e^{i\mathbf q\cdot\mathbf{R}}
\frac{\operatorname{erfc}\!\left[\gamma^{1/2}|\mathbf{R} + \mathbf{r}|\right]}
{|\mathbf{R} + \mathbf{r}|},\\
&S_{\rm rec.}(\mathbf q, \mathbf{r})=
\frac{\pi}{A_{\rm s.c.}}
\sum_{\mathbf G \neq \mathbf{q}}
\frac{e^{i(\mathbf G-\mathbf q)\cdot \boldsymbol{\rho}}}
{|\mathbf G-\mathbf q|}
     \left[e^{-|\mathbf{G} - \mathbf q||z|}
\operatorname{erfC}\!\left(\frac{|\mathbf{G} - \mathbf q|}{2\gamma^{1/2}}-\gamma^{1/2}|z|\right) +
e^{|\mathbf{G} - \mathbf q||z|}
\operatorname{erfC}\!\left(\frac{|\mathbf{G} - \mathbf q|}{2\gamma^{1/2}}+\gamma^{1/2}|z|\right)\right]\nonumber\\
&\hspace{2cm} -\frac{2\pi}{A_{\rm s.c.}}\left( z \text{erf}(\gamma^{1/2} z) + \frac{1}{\sqrt{\pi \gamma}} e^{- \gamma z^2} \right) \delta_{\mathbf{q}, \mathbf{0}} - \frac{2\sqrt{\gamma}}{\sqrt{\pi}} \delta_{\mathbf{r}, \mathbf{0}}
\end{align}
We removed singular contribution at $\mathbf{q} = \mathbf{0}$ which are canceled by neutralizing planes at $z = \pm t_s/2$. 
From these expressions, classical energy per particle for a basis $\mathbf{r}_{\alpha}$, is given in Ewald summation as 
\begin{align*}
    u  = \frac{1}{2N} \sum_{\alpha \ell \beta \ell'} \sideset{}{'}\sum_{\mathbf{R}} \ \frac{1}{|\mathbf{R} + \mathbf{r}_{\alpha \ell} -\mathbf{r}_{\beta \ell'}|}
     + \frac{n_{\rm 2D}\pi t_s}{2} = \frac{1}{2N} \sum_{\alpha \ell \beta \ell'} S(\mathbf{0}, \mathbf{r}_{\alpha \ell} -\mathbf{r}_{\beta \ell'}) + \frac{n_{\rm 2D}\pi t_s}{2}. 
\end{align*}
where the second term combines electron-background and background-background interaction.

To write the Hessian contribution in direct space, we introduce 
\begin{align}
f(r) &= \frac{\operatorname{erfc}\left[\gamma^{1/2}r\right]}{r}, \quad
f'(r) = -\frac{\operatorname{erfc}\left[\gamma^{1/2}r\right]}{r^2}
- \frac{2\gamma^{1/2}}{\sqrt{\pi} r} e^{- \gamma r^2}, \quad
f''(r) = 2\frac{\operatorname{erfc}\left[\gamma^{1/2}r\right]}{r^3}
+ e^{- \gamma r^2} \left(\frac{4\gamma^{1/2}}{\sqrt{\pi} r^2}
+ \frac{4\gamma^{3/2}}{\sqrt{\pi}}\right) 
\end{align}
and get
\begin{align}
\partial_{\mu} \partial_{\nu} S_{\rm dir.}(\mathbf{q}, \mathbf{r})
=
\sideset{}{'} \sum_{\mathbf{R}} e^{i\mathbf q\cdot\mathbf R}
\left(
\frac{(\mathbf{R} + \mathbf{r})_{\mu}(\mathbf{R} + \mathbf{r})_{\nu}}{|\mathbf{R} + \mathbf{r}|^2}
\left(- \frac{f'(|\mathbf{R} + \mathbf{r}|)}{|\mathbf{R} + \mathbf{r}|} + f''(|\mathbf{R} + \mathbf{r}|)\right)
+ f'(|\mathbf{R} + \mathbf{r}|) \frac{\delta_{\mu \nu}}{|\mathbf{R} + \mathbf{r}|}
\right)
\end{align}
where $\partial_{\mu} = \frac{\partial }{\partial \mathbf{r}_\mu}$ for $\mu = x, y$ in-plane directions. In parallel, the Hessian reciprocal space contribution can be written as 
\begin{align}
&\partial_{\mu} \partial_{\nu} S_{\rm rec.}(\mathbf{q}, \mathbf{r}) =
-\frac{\pi}{A_{\rm s.c.}}
\sum_{\mathbf G \neq \mathbf{q}}
(\mathbf G - \mathbf q)_{\mu} (\mathbf G - \mathbf q)_{\nu}
\frac{e^{i(\mathbf G - \mathbf q)\cdot \boldsymbol{\rho}}}
{|\mathbf G - \mathbf q|}
\, \bigg[e^{-|\mathbf{G} - \mathbf q||z|}
\operatorname{erfC}\!\left(\frac{|\mathbf{G} - \mathbf q|}{2\gamma^{1/2}}-\gamma^{1/2}|z|\right) \nonumber \\
& \hspace{8.6cm}+
e^{|\mathbf{G} - \mathbf q||z|}
\operatorname{erfC}\!\left(\frac{|\mathbf{G} - \mathbf q|}{2\gamma^{1/2}}+\gamma^{1/2}|z|\right)\bigg]. 
\end{align}
The elements of the dynamical matrix \cite{gonzeDynamicalMatricesBorn1997} are given by
\begin{align}
[\mathbf{D}^{\ell\ell'}_{\alpha\beta}(\mathbf q)]_{\mu\nu}
&=
-\partial_\mu\partial_\nu S(\mathbf q, \mathbf{r}_{\alpha \ell} -\mathbf{r}_{\beta \ell'})
+\delta_{\alpha \beta} \delta_{\ell \ell'}\displaystyle\sum_{\beta' \ell''}\partial_\mu\partial_\nu\, S(\mathbf 0, \mathbf{r}_{\alpha \ell} -\mathbf{r}_{\beta' \ell''}).
\end{align}  


\section{General Scalings}

\label{sec:angle_dep}

The mechanism underlying the angle dependence of our results is illustrated by examining the interlayer Coulomb interaction energy per particle $u_{\rm inter}$ for a basis $\mathbf{r}_{\alpha, \ell} = (\boldsymbol{\rho}_{\alpha, \ell}, z_{\alpha, \ell})$. In momentum space, this energy can be written as
\begin{align}
    u_{\rm inter}
    &=  \frac{1}{2N} \sum_{\alpha\beta,  \ell \neq \ell'} \sum_{\mathbf{R}} \ \frac{1}{|\mathbf{R} + \mathbf{r}_{\alpha \ell} -\mathbf{r}_{\beta \ell'}|}
     + \frac{n_{\rm 2D}\pi t_s}{2} \nonumber\\
    &= \frac{1}{2N} \sum_{\alpha \beta, \ell \neq \ell'}  \left(- \frac{2\pi}{A_{\rm s.c.}} t_s + \frac{2\pi}{A_{\rm s.c.}} \sum_{\mathbf{G} \neq \mathbf{0}} \frac{e^{- |\mathbf{G}| t_s}}{|\mathbf{G}|} e^{i \mathbf{G }\cdot (\mathbf{r}_{\alpha, \ell} - \mathbf{r}_{\beta, \ell'})} \right) 
     + \frac{n_{\rm 2D}\pi t_s}{2} \nonumber\\
    &=  \frac{\pi n_{\rm 2D}}{2} \sum_{\mathbf{G} \neq \mathbf{0}} \frac{e^{- |\mathbf{G}| t_s}}{|\mathbf{G}|}  \mathrm{Re}[\rho_1(\mathbf{G})\rho_2(-\mathbf{G})]  \label{eq:uinter}
\end{align}
where $\mathbf{G}$ ranges over the reciprocal lattice of the superlattice and $\rho_\ell(\mathbf{G}) = \frac{2}{N}\sum_{\alpha} e^{i \mathbf{G} \cdot \mathbf{r}_{\alpha, \ell}}$. The interlayer interaction is controlled by the weight $\mathrm{Re}[\rho_1(\mathbf{G})\rho_2(-\mathbf{G})]$ that selects the contributing $\mathbf{G}$. For a honeycomb stacking, the relative shift $\mathbf{s}$ in the origin of the two layers gives $\mathrm{Re}[\rho_1(\mathbf{G})\rho_2(-\mathbf{G})] = \cos(\mathbf{G} \cdot \mathbf{s}) < 0$ and $\mathbf{G}$ runs over the monolayer reciprocal lattice shared by layers $1$ and $2$. This negative contribution reflects the tendency of layers to relax towards $AB$ stacking to lower their energy relative to the decoupled limit.

\begin{figure}
    \centering
    \includegraphics[width=0.95\linewidth]{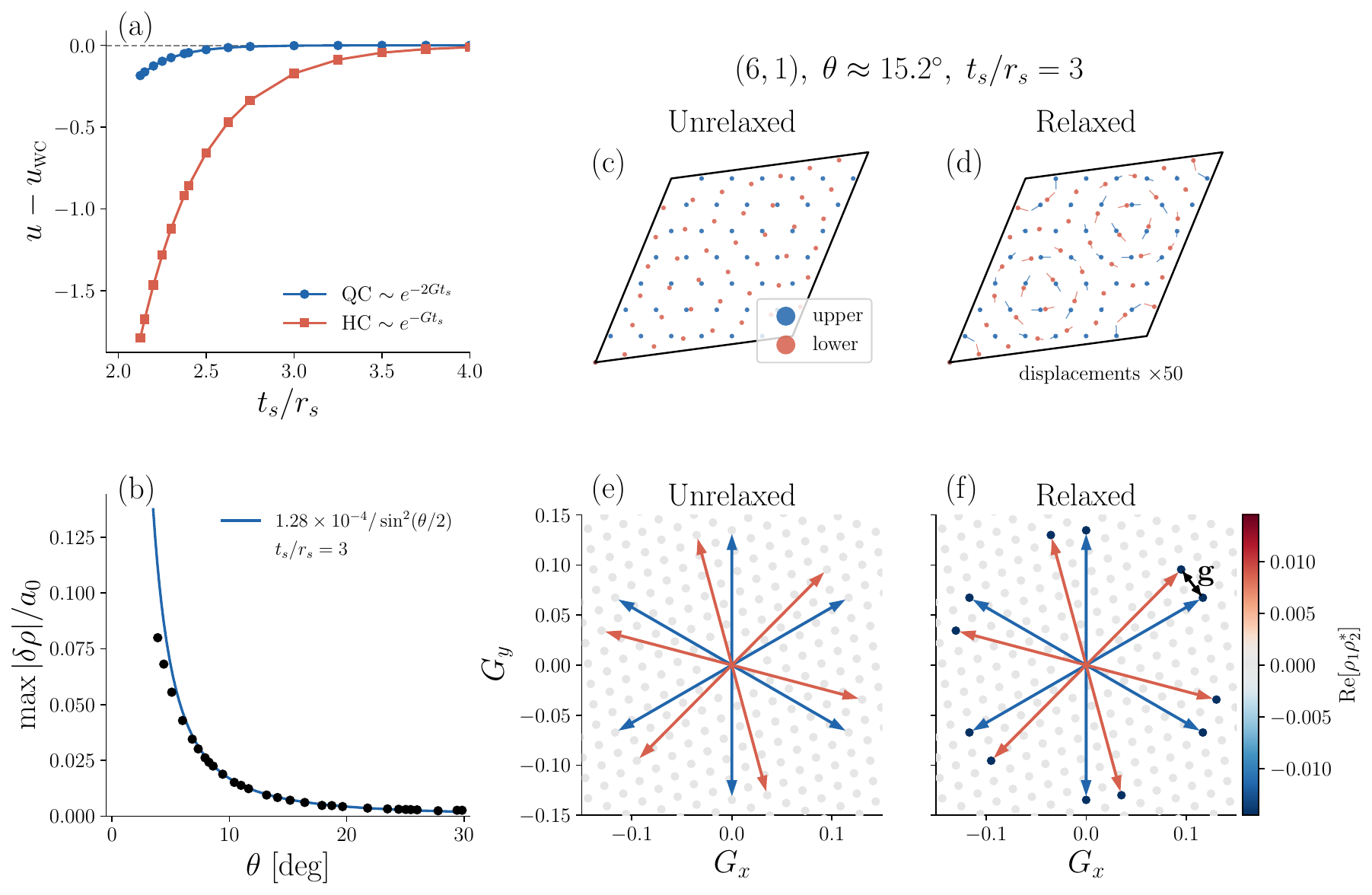}
    \caption{(a) Comparison with $u_{\rm \scriptscriptstyle WC}$ ($10^{-3} E_{\rm H} \cdot r_s^{-1}$) of the classical coulomb energy of the bilayer against $t_s/r_s$ for $\theta \approx 29.84^\circ$ and honeycomb stacking. (b) Magnitude of the maximal displacement $\delta \boldsymbol{\rho}$ from relaxation as a function of $\theta$ for $t_s/r_s =3$. A fit for the general angle dependence of Eq. \ref{eq:angle_scaling} is shown in blue. (c, d) Initial and relaxed configurations for commensurate twist $(6, 1)$. Relaxation displacement is scaled by $50$ and indicated by a line joining initial position and plot point of the final position. (d, e) Interlayer structure factor $\text{Re}[\rho_1(\mathbf{G})\rho_2(-\mathbf{G})]$ from Eq. \ref{eq:uinter} for the unrelaxed and relaxed configuration highlighting the role of relaxation in the formation of Bragg peaks at non-trivial $\mathbf{G}$. Points $\mathbf{G}$ of the reciprocal superlattice are shown in gray. Blue (resp. red) arrows represent the first shell of reciprocal lattice vectors for the lower (resp. upper) layer. A minimal moiré vector $\mathbf{g}$ connecting the reciprocal lattices of both layers is shown with a double arrow}
    \label{fig:relax}
\end{figure}

We now contrast the behavior of the unrelaxed $\rho^U_{\ell}$ and relaxed $\rho^R_{\ell}$ density Fourier components for twisted configuration. For a general commensurate twist \textit{without relaxation}, $\mathbf{r}_{\alpha \ell} = \mathbf{r}_{\alpha \ell}^U$ is a rigid triangular lattice in each layer. The resulting $\text{Re}[\rho_1^U(\mathbf{G})\rho_2^U(-\mathbf{G})]$ only has peaks at coincidence points of the triangular reciprocal lattices of layers $1$ and $2$, which form the \textit{super- reciprocal lattice} discussed at the end of Sec.\ \ref{sec:CS}. Before relaxation, the shortest non-zero coinciding reciprocal vectors for generic twist have modulus $|\mathbf{G}|^2 = G^2 N/2$ and their contribution are exponentially suppressed with $\exp (- G t_s\sqrt{N/2} )$, making them negligible.

In the relaxed case, $\mathbf{r}_{\alpha \ell}^R = \mathbf{r}_{\alpha \ell}^U + \delta \mathbf{r}_{\ell}(\mathbf{r}_{\alpha \ell}^U)$ where $\delta \mathbf{r}_{\ell}$ is the displacement at $\mathbf{r}_{\alpha \ell}^U$ caused by relaxation of layer $\ell$. This displacement repeats with the periodicity of the superlattice and can therefore be expanded in associated Fourier mode $\mathbf{g}$ as $\delta \mathbf{r}_{\ell}(\mathbf{r}) = \sum_{\mathbf{g}} \boldsymbol{a}_{\ell, \mathbf{g}} e^{i \mathbf{g} \cdot \boldsymbol{\rho}}$. We take $\mathbf{a}_{\ell, \mathbf{0}} = \mathbf{0}$ by absorbing uniform relative shifts of the layers into the unrelaxed configuration. For small $\delta \mathbf{r}_\ell$, it follows that the Fourier components of the relaxed density are
\begin{align}
    \rho_\ell^R(\mathbf{G}) &= \sum_{\alpha} e^{i \mathbf{G} \cdot \mathbf{r}_{\alpha \ell}^U} (1 +  i \mathbf{G}\cdot \delta \mathbf{r}_{\ell}( \mathbf{r}_{\alpha \ell}^U)) + O(\delta \mathbf{r}^2_{\ell})
    = \rho_\ell^U(\mathbf{G})  +  i \mathbf{G}\cdot \sum_{\mathbf{g}} \rho_{\ell}^U(\mathbf{G} + \mathbf{g})  \mathbf{a}_{\ell, \mathbf{g}} + O(\delta \mathbf{r}^2_{\ell}).
\end{align}
Compared to $\rho_\ell^U$, the relaxed Fourier components $\rho_\ell^R$ feature a richer set of peaks: layer $\ell$ acquires contributions from reciprocal vectors $\mathbf{G}$ of layer $\ell'$, provided $\mathbf{g} + \mathbf{G}$ still belongs to the reciprocal lattice of layer $\ell$. In particular, we can simultaneously have 
$\mathbf{G}$ in the reciprocal lattice of layer $2$, so that $\rho_2^R(-\mathbf{G}) \neq 0$, and $\mathbf{g} + \mathbf{G}$ in that of layer $1$, so that $\rho_1^R(\mathbf{G} + \mathbf{g}) \neq 0$.

To first order in $\delta \mathbf{r}_\ell$, the interlayer interaction for the component $\mathbf{G}$ is then proportional to
\begin{align}
    \text{Re}[\rho_1^R(\mathbf{G})\rho_2^R(-\mathbf{G})] &=\text{Re}[\rho_1^U(\mathbf{G})\rho_2^U(-\mathbf{G})] \nonumber \\
    &\quad \quad + \text{Re} \left[i \mathbf{G}\cdot \sum_{\mathbf{g}} \left(\rho_{1}^U(\mathbf{G} + \mathbf{g}) \rho_{2}^U(-\mathbf{G}) \mathbf{a}_{1, \mathbf{g}}  - \rho_{1}^U(\mathbf{G})\rho_{2}^U(-\mathbf{G} + \mathbf{g}) \mathbf{a}_{2, \mathbf{g}} \right)\right] + O(\delta \mathbf{r}_\ell^2) \label{eq:RE}
\end{align}
 The cross terms in Eq.~\ref{eq:RE} typically dominate over the negligible unrelaxed contribution $\text{Re}[\rho_1^U(\mathbf{G})\rho_2^U(-\mathbf{G})]$, and are maximal for relaxation wavevectors $\mathbf{g} = \mathbf{G}_1 - \mathbf{G}_2$ with $|\mathbf{g}| = 2G\sin(\theta/2) \sim L_M^{-1}$ setting the inverse moiré length scale.
In Fig.~\ref{tab:placeholder} (c-f), we show an example of the effect of relaxation on the contribution of $\mathbf{G}_\ell$ to $\text{Re}[\rho_1(\mathbf{G})\rho_2(-\mathbf{G})]$. As expected, unrelaxed $\text{Re}[\rho_1^U(\mathbf{G})\rho_2^U(-\mathbf{G})]$ (e) does not feature peaks at small non-zero $\mathbf{G}_{\ell}$  and (f) shows how these peaks are developed through relaxation.

While $\text{Re}[\rho_1^R(\mathbf{G})\rho_2^R(-\mathbf{G})] \neq 0$ is realized in principle for every $\mathbf{G}_\ell$, exponential suppression $e^{-|\mathbf{G}|t_s}$ selects the minimal non-zero $|\mathbf{G}| = G$ shell as the dominant contribution. Another essential factor influencing the contribution of $\mathbf{G}_{\ell}$ is the Fourier component $\mathbf{a}_{\ell, \mathbf{g}}$. In what follows, we obtain the scaling of $\mathbf{a}_{\ell, \mathbf{g}}$ with well thickness and its angle dependence which carries over to our general results for classical and zero point energy.

We argued in the main text that, for sufficiently large $t_s/r_s$ at a finite twist, relaxation contributes in a perturbative way. In this regime, we found $\delta \mathbf{r} \sim e^{-G t_s} L_M^{2}/a$ and we now provide a more precise version of this result. We focus the displacement $\delta \mathbf{r}_1$ induced by relaxation. 

\begin{figure}
    \centering
    \includegraphics[width=0.75\linewidth]{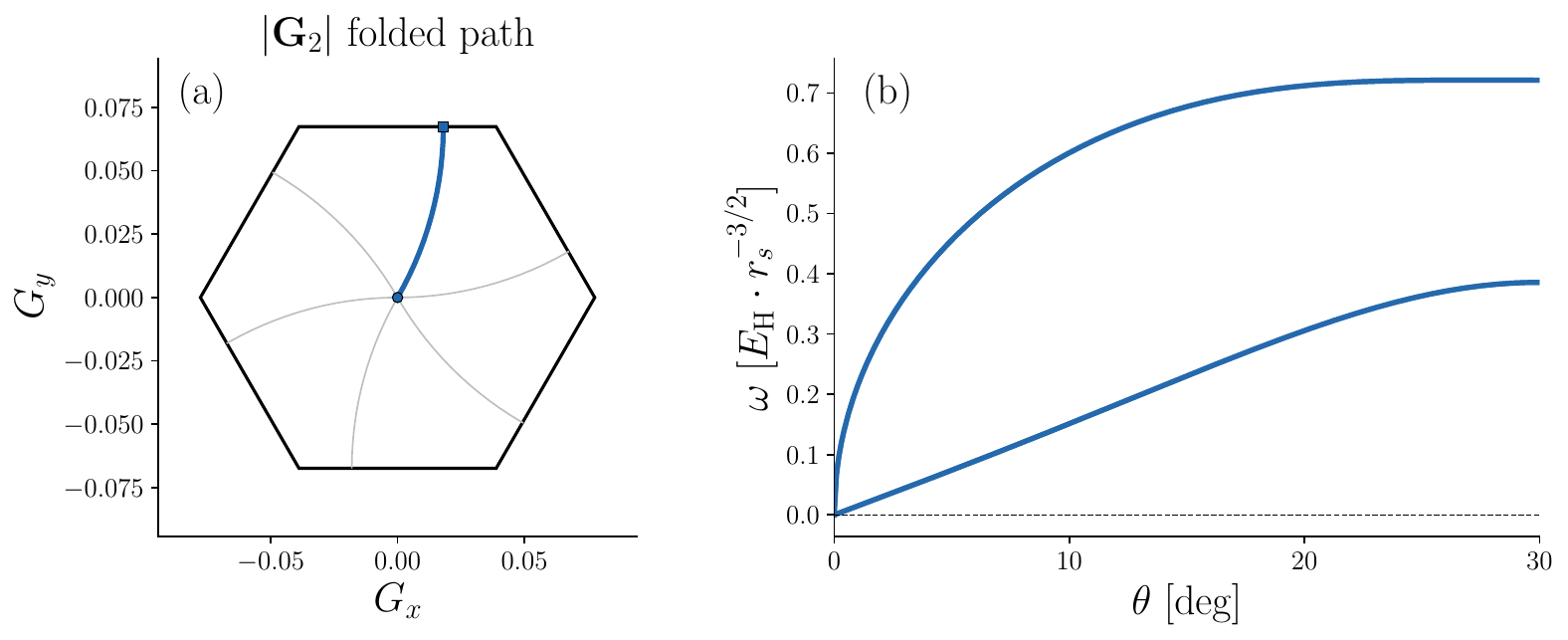}
    \caption{(a) Path traced by the  smallest non-zero $\mathbf{G}_2$ folded back to the Brillouin zone of layer $1$ as angle $\theta$ varies from $0$ to $30^\circ$, (b) Cut of the band structure of a monolayer Wigner crystal along the folded $\mathbf{G}_2$ path.}
    \label{fig:Gpath}
\end{figure}

We start by expanding $\delta \mathbf{r}_1$ against the normal modes of a triangular Wigner crystal in layer $1$:
\begin{align}
    \delta \mathbf{r}_{1}(\mathbf{r}) = \sum_{\mathbf{q}} (a_{1, \mathbf{q}}^T \mathbf{e}^T_{1, \mathbf{q}} + a_{1, \mathbf{q}}^L \mathbf{e}^L_{1, \mathbf{q}}) e^{i \mathbf{q} \cdot \boldsymbol{\rho}}
\end{align}
where $\omega_T(\mathbf{q})$ and $\omega_L(\mathbf{q})$ are the acoustic phonon and longitudinal plasmon frequencies with respective polarizations $\mathbf{e}^{T} = \hat{z} \times \hat{q}$ and $\mathbf{e}^{L} = \hat{q}$.  
Here $\mathbf{q}$ belongs to the Brillouin zone of monolayer Wigner crystal. However, as stated above, we suppose that relaxation repeats with the periodicity of the superlattice and only specific $\mathbf{q} = \mathbf{g}$ in the reciprocal superlattice are allowed. With this in mind, we write the forces acting on the electron with an unrelaxed position $\mathbf{r}$ with displacement $\delta \mathbf{r}_1$. The elastic intra-layer restoring force is  
\begin{align}
    \mathbf{F}_{\rm intra}(\mathbf{r}) = -\sum_{\mathbf{q}} (\omega_T(\mathbf{q})^2 a_{1, \mathbf{q}}^T \mathbf{e}^T_{1, \mathbf{q}} + \omega_L(\mathbf{q})^2 a_{1, \mathbf{q}}^L \mathbf{e}^L_{1, \mathbf{q}}) e^{i \mathbf{q} \cdot \boldsymbol{\rho}}. 
\end{align}
The force applied by layer $2$ depends on the relaxation of this layer. However, the dominant contribution to this force is captured by the potential sourced by the unrelaxed configuration of layer $2$, which only has Fourier components at reciprocal lattice vectors $\mathbf{G}_2$ of the rigid triangular layer $2$. Neglecting exponentially supressed shells with $|\mathbf{G}_2| > G$, the force associated with the rigid configuration is 
\begin{align}
    \mathbf{F}_{\rm inter}(\mathbf{r}) &= -i \pi n_{\rm 2D} \sum_{|\mathbf{G}_2| = G} \frac{\mathbf{G}_2}{G}  e^{i \mathbf{G}_2 \cdot \boldsymbol{\rho}} e^{-G t_s} 
\end{align}
To make contact with the intra layer force, we express $\mathbf{F}_{\rm inter}$ in the basis of plane waves on layer $1$ by translating $\mathbf{G}_2$ vectors back to the Brillouin zone of layer $1$. For the shell with $|\mathbf{G}_2| = G$, this process yields components at $\mathbf{g}$ moiré vectors with $|\mathbf{g}| = 2G\sin(\theta/2)$ since $e^{i \mathbf{G}_2 \cdot \boldsymbol{\rho}} = e^{i (\mathbf{g} + \mathbf{G}_1) \cdot \boldsymbol{\rho}} =  e^{i\mathbf{g} \cdot \boldsymbol{\rho}}$ for $\mathbf{r}$ in the rigid triangular lattice of layer $1$. The path of folded $\mathbf{G}_2$ as a function of $\theta$ is given Fig. \ref{fig:Gpath} (a). We have 
\begin{align*}
     \mathbf{F}_{\rm inter}(\mathbf{r}) = -i \pi n_{\rm 2D} \sum_{|\mathbf{g}| = 2G\sin(\theta/2)} \frac{1}{G}  e^{i \mathbf{g}  \cdot \boldsymbol{\rho}} e^{-G t_s} ((\mathbf{G}_2 \cdot \mathbf{e}^T_{\mathbf{g}}) \mathbf{e}^T_{\mathbf{g}} + (\mathbf{G}_2 \cdot \mathbf{e}^L_{\mathbf{g}}) \mathbf{e}^L_{\mathbf{g}}). 
\end{align*}
Balancing the forces, Fourier component by Fourier component, we have 
\begin{align}
    &a_{1, \mathbf{g}}^T = \frac{i \pi n_{\rm 2D}}{\omega_T(\mathbf{g})^2} e^{- G t_s} \cos(\theta/2) \quad \text{and} \quad a_{1, \mathbf{g}}^L = -\frac{i \pi n_{\rm 2D}}{\omega_L(\mathbf{g})^2} e^{- G t_s} \sin(\theta/2)  \label{eq:ag}
\end{align}
indicating that the effect of relaxation is set by frequencies of the $T, L$ modes evaluated (see Fig. \ref{fig:Gpath} (b)) over the folded $\mathbf{G}_2$ path of Fig. \ref{fig:Gpath} (a). For angles approaching $30^\circ$, $\mathbf{g}$ move towards the Brillouin zone boundary. There, Fig. \ref{fig:Gpath} (b) shows that the $\mathbf{g}$ dependence of frequencies saturates to satisfy $\partial \omega/\partial q_\perp = 0$ at the boundary. It follows that maximal frequencies and minimal relaxation occur at $30^\circ$ with a weaker angle dependence close to $30^\circ$. 

Looking at the denominators of Eq. \ref{eq:ag}, we observe that the dominant components $a_{1, \mathbf{g}}$ are associated with minimal transverse mode frequencies $\omega_T \approx c_T |\mathbf{q}|$ where $c_T$ is the speed of sound. The corresponding displacement is 
\begin{align}
\delta \mathbf{r}_1(\mathbf{r}) \approx i \sum_{|\mathbf{g}| = 2G\sin(\theta/2)}\frac{\pi n_{\rm 2D}}{c_T^2 |\mathbf{g}|^2} e^{- G t_s} \cos(\theta/2) e^{i\mathbf{g} \cdot \boldsymbol{\rho}} \hat{z} \times \hat{g} \sim \frac{L_M^2}{a} e^{- G t_s}. \label{eq:angle_scaling}
\end{align}

Carrying the same analysis and approximations for layer $2$, we find $\delta \mathbf{r}_{2}(\mathbf{r}) = -\delta \mathbf{r}_{1}(\mathbf{r})$.
The general small angle scaling of Eq.\eqref{eq:angle_scaling} is verified numerically in Fig. \ref{fig:relax} (b) and is consistent with the analogous results in the context of twisted bilayer graphene \cite{namLatticeRelaxationEnergy2017, ezziAnalyticalModelAtomic2024}. 

With the displacement induced by relaxation, we can find the equillirbium Coulomb energy per particle by substitution into Eq. \ref{eq:uinter} and into the elastic expansion of intralayer energy per particle
\begin{align*}
    u_{\rm intra} = u_{\rm \scriptscriptstyle WC} + \frac{1}{4} \sum_{\ell} \sum_{\mathbf{g}} (\omega_{T}(q)^2 |a_{\ell, \mathbf{g}}^T|^2 + \omega_{L}(q)^2 |a_{\ell, \mathbf{g}}^L|^2)
\end{align*}
From Eq.~\ref{eq:RE}, the exponentially supressed interlayer interaction combines with the exponential supression of relaxation induced Bragg peaks $\propto \mathbf{a}_{\ell, \mathbf{g}}$ to yield $u_{\rm inter} \sim L_M^2 e^{-2Gt_s}/a$. The intralayer interaction relative to $u_{\rm \scriptscriptstyle WC}$ contributes at $O(\delta \mathbf{r}_{\ell}^2)$ with the same scaling $u_{\rm intra} \sim \omega(\mathbf{g})^2 \delta \mathbf{r}^2_\ell \sim L_M^2 e^{-2 Gt_s}/a$. It follows that $u-u_{\rm \scriptscriptstyle WC}$ globally features this shared scaling. We verify this claim in Fig. \ref{fig:relax} (a). We compare with honeycomb stacking for which there is no relaxation and where the rigid interlayer interaction directly sets $u \sim e^{- G t_s}/a$.

\section{Soft modes}
\label{sec:soft}
In the limit of fully decoupled layers, the phonon spectrum consists of two copies of the monolayer Wigner crystal spectrum. Each layer $\ell$ supports a transverse acoustic mode of frequency $\omega_{\ell T}(\mathbf{q})$ and a longitudinal plasmon mode of frequency $\omega_{\ell L}(\mathbf{q})$, defined within its own Brillouin zone.

As $t_s/r_s$ is reduced, the interlayer interaction grows and couples phonons from the two layers. This coupling breaks the independent translation symmetry of each layer down to the periodicity of the commensurate supercell, and the natural object in reciprocal space becomes the mini-Brillouin zone (mBZ) of the superlattice, constructed from $\mathbf{B}_1$ and $\mathbf{B}_2$ (see Section~\ref{sec:CS}). Different $\mathbf{q}$ sectors of the mBZ remain decoupled, and within each sector the leading effect of the interlayer interaction is to lift the degeneracy between modes of nearly identical frequency.

In aligned stackings such as honeycomb, modes from the two layers are degenerate at every $\mathbf{q}$, and the interlayer coupling lifts this degeneracy throughout the mBZ. At finite twist, the same mBZ wavevector probes different symmetry directions from the Brillouin of each layer, so the modes are already non-degenerate at the decoupled level -- except in the long-wavelength limit, where isotropy restores $\omega_{1m}(\mathbf{q}) = \omega_{2m}(\mathbf{q}) \equiv \omega_m(\mathbf{q})$, with $\omega_T = c_T|\mathbf{q}|$ and $\omega_L = \sqrt{\pi n_{\rm 2D} |\mathbf{q}|}$ ($c_T$ the transverse speed of sound). The mismatch is maximal at large wavelengths for $30^\circ$, where the two Brillouin zones are maximally rotated relative to one another and the dispersion minima of one layer fall on maxima of the other. The mixing of the two modes is weak, and only the small isotropic region near $\mathbf{q} = \mathbf{0}$ feels the effect of interlayer coupling strongly.

We now compute the degeneracy lifting for the soft transverse and longitudinal modes near $\mathbf{q} = \mathbf{0}$, where the two layers contribute degenerate copies and the interlayer interaction acts most strongly. The honeycomb and quasicrystal stackings are compared in Fig.~\ref{fig:Band_DOS}$\,(b,c)$. The relevant normal modes carry a wavevector $\mathbf{q}$ that lies simultaneously within the Brillouin zone of each individual layer and within the mBZ of the supercell. In the Brillouin zone of layer $\ell$, mode $m = T, L$ takes the form $\mathbf{e}_m(\mathbf{q})\,e^{i\mathbf{q}\cdot\mathbf{R}}$, where $\mathbf{e}_m$ is the polarization of the single electron in the monolayer basis and $\mathbf{R}$ runs over monolayer sites. To make contact with modes of the other layer, we re-express this mode in the mBZ as $\mathbf{e}_{m,\alpha,\ell}(\mathbf{q}) \, e^{i\mathbf{q}\cdot\mathbf{R}'}$ with
\begin{align}
    \mathbf{e}_{m,\alpha,\ell}(\mathbf{q}) = (N/2)^{-1/2}\,\mathbf{e}_m(\mathbf{q})\,e^{i\mathbf{q}\cdot\mathbf{r}_{\alpha,\ell}},
\end{align}
and $\mathbf{R}'$ now runs over superlattice vectors. Projecting the dynamical matrix onto these four modes -- two polarizations $m = T, L$ for each of the two layers $\ell = 1, 2$ -- yields a $4 \times 4$ matrix with components
\begin{align}
    D^{\ell \ell'}_{m m'}(\mathbf{q}) &= \sum_{\alpha \beta} \left(\mathbf{e}_{m', \beta, \ell'}\right)^\dagger \mathbf{D}^{\ell \ell'}_{\alpha \beta}(\mathbf{q})\, \mathbf{e}_{m, \alpha, \ell} \nonumber \\
    &= \sum_{\alpha \beta, \mu \nu}\bigg(-\partial_\mu\partial_\nu S(\mathbf{q}, \mathbf{r}_{\alpha\ell} - \mathbf{r}_{\beta\ell'}) + \delta_{\alpha\beta}\delta_{\ell\ell'}\sum_{\beta'\ell''}\partial_\mu\partial_\nu S(\mathbf{0}, \mathbf{r}_{\alpha\ell} - \mathbf{r}_{\beta'\ell''})\bigg)(e^\nu_{m', \beta, \ell'})^\star\, e^\mu_{m, \alpha, \ell},
\end{align}
where,
\begin{align}
    \partial_\mu\partial_\nu S(\mathbf{q}, \mathbf{r}) = -\frac{2\pi}{A_{\rm s.c.}}\sum_{\mathbf{G} \neq \mathbf{q}} (\mathbf{G} - \mathbf{q})_\mu (\mathbf{G} - \mathbf{q})_\nu\, \frac{e^{i(\mathbf{G} - \mathbf{q})\cdot\boldsymbol{\rho}}}{|\mathbf{G} - \mathbf{q}|}\, e^{-|\mathbf{G} - \mathbf{q}||z|}.
\end{align}
While the $L$ and $T$ sectors are exactly decoupled at $\mathbf{q} = \mathbf{0}$ by $C_3$ symmetry, away from $\Gamma$ the longitudinal plasmon $\omega_L \sim |\mathbf{q}|^{1/2}$ and the transverse mode $\omega_T \sim |\mathbf{q}|$ are split by a frequency ratio $\omega_L/\omega_T \to \infty$ as $\mathbf{q} \to \mathbf{0}$, suppressing $L$--$T$ mixing at small $|\mathbf{q}|$. The $4 \times 4$ matrix therefore reduces to two independent $2 \times 2$ blocks, one per polarization. Their components are
\begin{align}
    &D^{11}_{mm} = \omega_m(\mathbf{q})^2 + M_{m 1},\quad D^{22}_{mm} = \omega_m(\mathbf{q})^2 + M_{m 2}, \quad D^{21}_{mm} = (D^{12}_{mm})^\star = \Delta_m
\end{align}
with 
\begin{align}
M_{m\ell} &= \pi n_{\rm 2D} \sum_{\mathbf{G}\neq\mathbf{q}} \frac{|\rho_\ell^R(\mathbf{G})|^2 - |\rho_\ell^U(\mathbf{G})|^2}{|\mathbf{G}-\mathbf{q}|} \left(\mathbf{e}_{m}\cdot(\mathbf{G}-\mathbf{q})\right)^2 \nonumber\\
&\quad - \pi n_{\rm 2D}\sum_{\mathbf G \neq \mathbf{0}}
\left(\frac{|\rho_{\ell}^R(\mathbf{G})|^2 - |\rho_\ell^U(\mathbf{G})|^2}{|\mathbf G|} 
      + \frac{\rho_{\ell}^R(\mathbf{G})\rho_{\bar{\ell}}^R(-\mathbf{G})} 
{|\mathbf G|} 
     e^{-|\mathbf{G}| t_s}\right) \left(\mathbf{e}_{m}\cdot \mathbf{G}\right)^2,\\
\Delta_{m} &= \pi n_{\rm 2D} \sum_{\mathbf{G}\neq\mathbf{q}} \frac{\rho_1^R(\mathbf{G})\,\rho_2^R(-\mathbf{G})}{|\mathbf{G}-\mathbf{q}|} 
e^{-|\mathbf{G}-\mathbf{q}|t_s} 
\left(\mathbf{e}_m\cdot(\mathbf{G}-\mathbf{q})\right)^2
\end{align}
where the combination  
\begin{align}
    |\rho_\ell^R(\mathbf{G})|^2 - |\rho_\ell^U(\mathbf{G})|^2 =
    -2\sum_{\mathbf{g}} \text{Im}\left[\rho_{\ell}^U(\mathbf{G}+\mathbf{g})\rho_{\ell}^U(-\mathbf{G}) \mathbf{G}\cdot \mathbf{a}_{\ell, \mathbf{g}} \right] + O(\delta\mathbf{r}^2_{\ell}) \label{eq:rho11}
\end{align}
isolates the part of the intralayer Hessian driven purely by lattice relaxation to separate it from the unrelaxed layer $\omega_{m}(\mathbf{q})^2$ contributions. We observe that Eq.\eqref{eq:rho11} is non zero only when $\mathbf{g}$ connects two reciprocal lattice vectors of layer $\ell$. From the results of Sec. \ref{sec:angle_dep} for $\delta \mathbf{r}_{\ell}$, the physically relevant $\mathbf{g}$ are reciprocal lattice vectors $\mathbf{G}_{\ell'}$ of the opposite layer, folded into the Brillouin zone of layer $\ell$; those that fold onto a vector $\mathbf{G}_\ell$ have magnitude $|\mathbf{G}_{\ell'}| \sim \sqrt{N}$ and have negligible effect. It follows that Eq. \eqref{eq:rho11} is $O(\delta \mathbf{r}_\ell^2)$ and has the same $t_s/r_s$ scaling as the remaining terms in $M_{m\ell}$ and $\Delta_m$. Keeping the minimal shell $|\mathbf{G}| = G$, we have the following general scalings
\begin{align}
\Delta_T &= O(r), \quad \Delta_L = \frac{\pi n_{\rm 2D}}{2} 
|\mathbf q|e^{-|\mathbf{q}| t_s}
 + O(r),\quad
M_{T} = O(r), \quad 
M_{L} = O(r).
\end{align}
where $r = e^{-2 G t_s}$ for twists in the perturbative regime and $r = e^{- G t_s}$ for honeycomb stacking. 

The degeneracy lifted modes with frequencies $\Omega_m^{\pm}(\mathbf{q})$ for $D_{mm}^{\ell \ell'}$ are 
\begin{align}
    \Omega_m^\pm(\mathbf{q}) = \left[\omega_m(\mathbf{q})^2 + \frac{1}{2}(M_{m1} + M_{m2}) \pm \frac{1}{2}\sqrt{(M_{m1} - M_{m2})^2 + 4|\Delta_{m}|^2}\right]^{1/2}.
\end{align}

While in-phase modes are protected by translation symmetry, the gap $\Omega^{\pm}(\mathbf{0})$ developed at $\mathbf{q} = \mathbf{0}$ for out of phase mode $m$ is given by 
\begin{align*}
    (\Omega^{\pm}(\mathbf{0}))^2= -2\Delta_{m} = -2\pi n_{\rm 2D} \sum_{\mathbf{G}\neq\mathbf{0}} \frac{\rho_1^R(\mathbf{G})\,\rho_2^R(-\mathbf{G})}{|\mathbf{G}|} 
e^{-|\mathbf{G}|t_s} 
\left(\mathbf{e}_m\cdot \mathbf{G}\right)^2
\end{align*}
and scales like $\sqrt{-G^2 u_{\rm inter}}$ with $t_s/r_s$. 
In particular, this means that in the very large $t_s/r_s$ limit, the dominant contribution to zero point energy is $\sim e^{-G t_s/2}$ for honeycomb and $\sim L_M e^{-G t_s}$ for quasicrystal.

Of particular interest, the $\Omega_L^-(\mathbf{q})$ LO phason mode consists of out of phase longitudinal excitations in each layer. An over-density of one layer is paired with an under-density of the other, making the excitation neutral in plane with behavior governed by dipole-dipole interaction. For the quasicrystal, neglecting the gap, its $\mathbf{q} \to \mathbf{0}$ dispersion has linear behavior (as seen in Fig.\ \ref{fig:Band_DOS} (c)) 
\begin{align}
    \Omega_L^-(\mathbf{q})= \sqrt{ \pi n_{\rm 2D}  
|\mathbf q|(1-e^{-|\mathbf{q}| t_s})} = \sqrt{\pi n_{\rm 2D} t_s} |\mathbf{q}| + O(|\mathbf{q}|^{3/2}) 
\end{align} 
with a divergent speed of sound, as $t_s \to \infty$ restoring the behavior of the 2D monolayer plasmon. As a gapless neutral mode, the optical plasmon constitutes a 2D Pines demon \cite{pinesElectronInteractionSolids1956} in the quasicrystal spectrum. 

\begin{figure}
    \centering
    \includegraphics[width=0.9\linewidth]{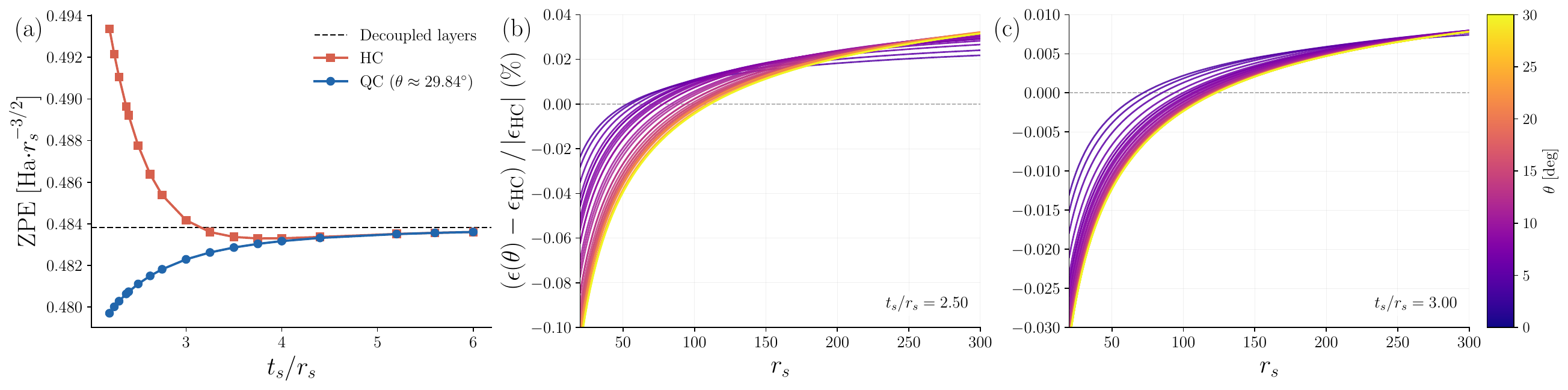}
    \caption{(a) Comparison of the ZPE as a function of $t_s/r_s$ for honeycomb and quasicrystal approximant at $\theta \approx 29.84^\circ$. Percentage energy difference with honeycomb for twist angle $\theta$ as a function of $r_s$ for $t_s/r_s = 2.5$ (b) and $t_s/r_s = 3$ (c)}
    \label{fig:relative_vs_rs}
\end{figure}

Both for honeycomb and twisted configurations in the large $t_s/r_s$ limit where the gap is small, we have the concavity inequality
\begin{align}
    \Omega_L^-(\mathbf{q}) + \Omega_L^+(\mathbf{q}) - 2 \omega_L(\mathbf{q}) \approx \sqrt{ \pi n_{\rm 2D}  
|\mathbf q|(1-e^{-|\mathbf{q}| t_s})}  + \sqrt{ \pi n_{\rm 2D}  
|\mathbf q|(1+e^{-|\mathbf{q}| t_s})} - 2 \sqrt{\pi n_{\rm 2D} |\mathbf{q}|} < 0 \label{eq:softer_decoupled}
\end{align}
and the softening of the out of phase longitudinal mode is such that the ZPE is globally lowered relative to decoupled layers. This behavior is seen in Fig.~\ref{fig:relative_vs_rs}$\,(a)$ for the large $t_s/r_s$ dependence of the ZPE of both honeycomb and quasicrystal approximant. While the lowering of ZPE relative to the decoupled limit persists to lower $t_s/r_s$ for the quasicrystal, the gap of the honeycomb stacking eventually becomes significant making the ZPE exceed the decoupled limit.






\section{Low $t_s/r_s$ considerations}
\label{sec:low_tr}

In this section, we comment on the low $t_s/r_s$ behavior of the bilayer.  Throughout, we have focused on the competition of various triangular stackings. Being the ground state of a monolayer Wigner crystal, triangular layers are favored when intralayer interaction dominates at large $t_s/r_s$. 
The mechanical stability of twisted triangular layers governs the allowed configurations of the system. Within the constraint of a given commensurate supercell, we find that twisted configuration become unstable and relax to honeycomb when interlayer interaction is sufficiently strong. In particular, the quasicrystal approximant at a $29.84^\circ$ twist relaxes back to honeycomb for $t_s/r_s \lesssim 1.95$ which is out of the large $t_s/r_s$ regions of interest of Fig. \ref{fig:ZPE}-\ref{fig:phase}.

\begin{figure}
    \centering
    \includegraphics[width=1\linewidth]{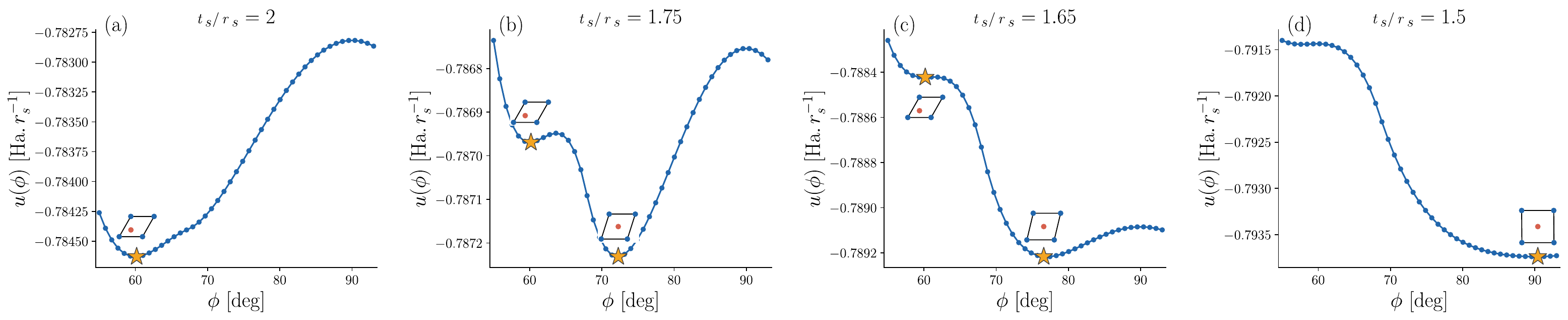}
    \caption{Classical energy landscape as a function of the unit cell shear angle $\phi$ of the untwisted bilayer Wigner crystal for different values of  $t_s/r_s$  \cite{goldoniStabilityDynamicalProperties1996}.}
    \label{fig:phi_sweep}
\end{figure}

We proceed to summarize the features of the classical landscape first studied in \cite{goldoniStabilityDynamicalProperties1996}. As interlayer interaction is increased, unit cells sheared at an angle $\phi$ between $\mathbf{a}_1^\ell$ and $\mathbf{a}_2^\ell$ start competing with the $\phi = 60^\circ$ honeycomb stacking. In Fig. \ref{fig:phi_sweep}, we obtain the minimal energy configuration for fixed unit cells with shear $\phi$ for different $t_s/r_s$. In panel $(a)$, we show a $\phi$ sweep at large $t_s/r_s$ highlighting the global classical energy minimum at the honeycomb stacking. Then panel (b) shows that from $t_s/r_s < 1.835$, the classical ground state discontinuously becomes a sheared configuration disconnected from the honeycomb configuration. The shear angle $\phi$ of this rhombic ground state is continuously increased as $t_s/r_s$ is further reduced (see panel $(c)$). Eventually, it reaches $\phi = 90^\circ$ at $t_s/r_s = 1.56$. Below this value, the staggered square ground state associated with this shear is the ground state of the system across all shears as can be seen in panel $(d)$. At even lower $t_s/r_s$, the bilayer crystal can lower its energy further by forming a staggered stacking of rectangles and finally, by becoming a monolayer triangular Wigner crystal. We point out that these phases are not accessible in a fully $3D$ quantum well where the bilayer collapses to a monolayer at small well thickness \cite{gaggioliElectronicCrystalsQuasicrystals2025}.

\end{document}